\documentclass[]{emulateapj}

\def\ni{\noindent}

\shorttitle{Cepheid P-L Relations in Sloan Magnitudes}
\shortauthors{Ngeow \& Kanbur}

\begin{document}

\title{Semi-Empirical Cepheid Period-Luminosity Relations in Sloan Magnitudes}

\author{C. Ngeow}
\affil{Department of Astronomy, University of Illinois, Urbana, IL 61801}

\and 

\author{S. M. Kanbur}
\affil{State University of New York at Oswego, Oswego, NY 13126}

\begin{abstract}

In this paper we derive semi-empirical Cepheid period-luminosity (P-L) relations in the Sloan $ugriz$ magnitudes by combining the observed $BVI$ mean magnitudes from the Large Magellanic Cloud Cepheids (LMC) and theoretical bolometric corrections. We also constructed empirical $gr$ band P-L relations, using the publicly available Johnson-Sloan photometric transformations, to be compared with our semi-empirical P-L relations. These two sets of P-L relations are consistent with each other.

\end{abstract}


\section{Introduction}

The Cepheid period-luminosity (P-L) relation is a fundamental tool in distance scale and stellar pulsation studies, which is traditionally studied in the standard Johnson-Cousin $BVI$ magnitude system. However, the magnitude system from the broad band filters employed in the Sloan Digital Sky Survey (SDSS, hereafter Sloan magnitudes) are becoming more popular in current and future studies. For example, observations with the Canada-France-Hawaii Telescope (CFHT), the Pan-STARRS\footnote{http://pan-starrs.ifa.hawaii.edu/public/home.html} \citep{kai04}, the Dark Energy Survey\footnote{https://www.darkenergysurvey.org/} \citep{des05}, the Large Synoptic Survey Telescope\footnote{http://www.lsst.org/lsst\_home.shtml} \citep[LSST,][]{tys02}, to name a few, will be done in the Sloan filters. In addition to wide field imaging, many of these surveys will include a time-domain component to detect asteroids and supernovae. Therefore, it is possible that a large number of variable stars, including Cepheids, will be observed and/or detected from these surveys. The purpose of this paper is to provide semi-empirical Cepheid P-L relations in the Sloan magnitudes for current and future Cepheid and distance scale studies (for example, the M33 Cepheids observed with CFHT as presented in \citealt{har06}). 

\section{Data, Method and Results}

Since the Large Magellanic Cloud (LMC) Cepheid P-L relation is extensively used in distance scale studies, we use the publicly available LMC Cepheid data from the OGLE \citep[Optical Gravitational Lensing Experiment,][]{uda99} database in this paper. The logarithmic periods, the extinction corrected $BVI$ band mean magnitudes and the extinction corrected $(B-V)$ colors for the fundamental mode LMC Cepheids were obtained from the OGLE database. We first remove Cepheids without a $(B-V)$ color in the OGLE sample. The remaining OGLE LMC Cepheids were then cross-correlated with the sample given in \citet{kan06} to remove outliers and ``bad'' Cepheids in our OGLE sample. Additional LMC Cepheids data are supplemented from table 1 of \citet{san04}. However we only include Cepheids with $0.4<\log P<1.8$ (where $P$ is pulsation period in days, see \citealt{kan06} for detailing why Cepheids are restricted to this range). Our final sample consists of $711$ LMC Cepheids with $BV$ band photometric data, $605$ of them having periods less than 10 days. From our sample, the fitted linear LMC P-L relations are:

\begin{eqnarray}
B & = & -2.370(\pm0.041)\log P + 17.346(\pm0.031),  \\
V & = & -2.715(\pm0.030)\log P + 17.048(\pm0.023), 
\end{eqnarray}

\ni with dispersion of $0.295$ and $0.222$ in $B$ and $V$ band, respectively. For completeness, we also include the $I$ band P-L relation from the $680$ Cepheids in our sample that have the $I$ band mean magnitudes (because the number of Cepheids with $(V-I)$ colors is less than those with $(B-V)$ colors in \citealt{san04} sample):

\begin{eqnarray}
I & = & -2.968(\pm0.021)\log P + 16.603(\pm0.016),  
\end{eqnarray}

\ni with dispersion of $0.147$. These P-L relations are in good agreement with the results presented by the OGLE team\footnote{ftp://sirius.astrouw.edu.pl/ogle/ogle2/var\_stars/lmc/cep/ \\ catalog/README.PL}, \citet{san04} and \citet{kan06}.

It is possible to derive semi-empirical P-L relations in the Sloan magnitudes by combining the LMC $BV$ Cepheid data with theoretical bolometric corrections ($BC$). In this paper we use the theoretical $BC$ from the Padova group \citep{gir02,gir04}, which are tabulated in grids with different metallicity, effective temperature ($T_{eff}$) and surface gravity ($\log [g]$). We first selected the $BC$ within the grids of $4000\mathrm{K} \leq T_{eff} \leq 7000\mathrm{K}$ and $0.0 \leq \log (g) \leq 2.5$, which are appropriate for classical Cepheids at mean light, from all of the available metallicity ($[M/H]=\{0.5,0.0,-0.5,-1.0,-1.5,-2.0,-2.5\} \mathrm{dex}$) given by the Padova group. To obtain the $BC$ for LMC Cepheids at $[M/H]=-0.3\mathrm{dex}$, these grids of $BC$ were then interpolated using the available metallicity. From the interpolated $BC$, we obtain a regression between $T_{eff}$, $\log (g)$ and $(B-V)$ color in the form of:

\begin{eqnarray}
\log T_{eff} & = & 3.8944 - 0.0058\log (g) \nonumber \\
& & - 0.2089 (B-V) + 0.0151(B-V)^2,
\end{eqnarray}

\ni which has a dispersion of $0.009$. Similarly, regressions for $BC$ in passband $\lambda$ takes the form of $BC_{\lambda}=\alpha_0+\alpha_1\log (g) + \alpha_2(\Delta T) + \alpha_3(\Delta T)^2 + \alpha_4(\Delta T)^3$, where $\Delta T=\log T_{eff}-3.772$. The coefficients in Johnson-Cousin $BVI$ and Sloan $ugriz$ passbands are given in Table \ref{tab1}. 

\begin{deluxetable}{lcccccc}
\tabletypesize{\scriptsize}
\tablecaption{Coefficients for bolometric corrections.\label{tab1}}
\tablewidth{0pt}
\tablehead{
\colhead{$\lambda$} & 
\colhead{$\alpha_0$} & 
\colhead{$\alpha_1$} & 
\colhead{$\alpha_2$} & 
\colhead{$\alpha_3$} & 
\colhead{$\alpha_4$} & 
\colhead{$\sigma$} }
\startdata
$B$  & $-0.5815$ & $+0.0242$ & $+7.4534$ & $-19.180$ & $+11.358$ & 0.057 \\ 
$V$  & $+0.0313$ & $-0.0111$ & $+2.0419$ & $-12.176$ & $+56.059$ & 0.012 \\ 
$I$  & $+0.6736$ & $-0.0115$ & $-2.2856$ & $-12.237$ & $+23.944$ & 0.009 \\
\multicolumn{7}{c}{} \\ 
$u$  & $-1.9787$ & $+0.1932$ & $+9.5140$ & $-60.418$ & $+12.036$ & 0.112 \\ 
$g$  & $-0.2506$ & $+0.0028$ & $+5.1822$ & $-14.651$ & $+44.885$ & 0.036 \\ 
$r$  & $+0.1306$ & $-0.0114$ & $+0.2299$ & $-12.250$ & $+39.899$ & 0.006 \\ 
$i$  & $+0.2009$ & $-0.0087$ & $-1.7603$ & $-13.445$ & $+26.189$ & 0.011 \\ 
$z$  & $+0.2462$ & $-0.0218$ & $-2.7303$ & $-7.8426$ & $+29.746$ & 0.004 
\enddata
\tablecomments{$\sigma$ is the dispersion of the regression.}
\end{deluxetable}

For each of the LMC Cepheids in our sample, the surface gravity can be estimated using $\log (g)=2.62-1.21\log P$ \citep{kov00,bea01}. Then the $T_{eff}$ and the $BC$ for our LMC Cepheids can be determined using the regressions we found previously. By combining the definitions of bolometric correction ($BC_{\lambda}=M_{\mathrm{bol}}-M_{\lambda}$) and distance modulus ($\mu=m_{\lambda}-M_{\lambda}$), we have $m_{\lambda}+BC_{\lambda}=M_{\mathrm{bol}}+\mu$. Then for each of the individual LMC Cepheids in our sample, $M_{\mathrm{bol}}+\mu$ is a constant and {\it independent of passband}. This quantity can be estimated using either $B+BC_B$ or $V+BC_V$ with available $BV$ mean magnitudes, or using either $V+BC_V$ or $I+BC_I$ with available $VI$ mean magnitudes. The mean difference between $B+BC_B$ and $V+BC_V$ for our LMC Cepheids is $0.005$mag. with a standard deviation of $0.007$mag., while the difference for $V+BC_V$ and $I+BC_I$ is $0.009$mag. with a larger standard deviation of $0.057$mag. For the $711$ LMC Cepheids in our sample, we adopt the averaged value of $M_{\mathrm{bol}}+\mu=[(B+BC_B)+(V+BC_V)+(I+BC_I)]/3$ if all three $BVI$ band data are available for a given Cepheid, else we use $M_{\mathrm{bol}}+\mu=[(B+BC_B)+(V+BC_V)]/2$ as the averaged value from $B$ and $V$ bands. Once the values of $M_{\mathrm{bol}}+\mu$ and the $BC$ in Sloan magnitudes are known for each Cepheid in the sample, the apparent Sloan magnitudes can be estimated by $m_{\lambda}=(M_{\mathrm{bol}}+\mu)-BC_{\lambda}$ and the P-L relations can be derived. 

\begin{figure}
\plotone{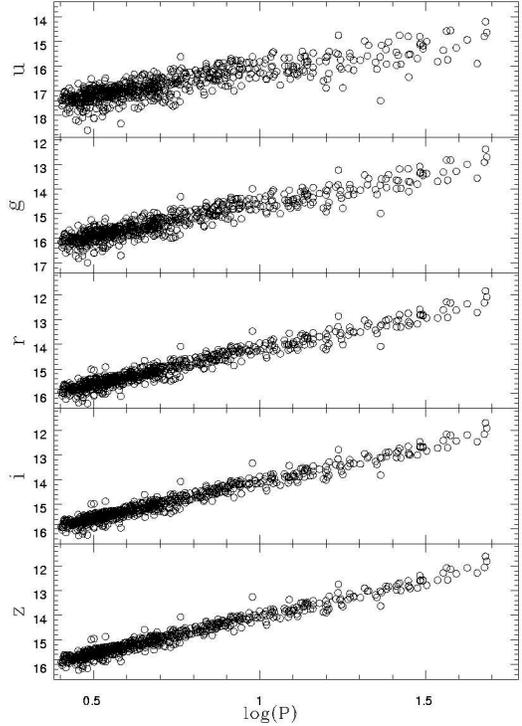}
\caption{The semi-empirical P-L relations in Sloan magnitudes using the method described in Section 2.\label{pl}}
\end{figure}

To test the above semi-empirical approach, we assume that only the LMC $B$ band data are available for the $711$ Cepheids and derive the semi-empirical $V$ band P-L relation. Using $M_{\mathrm{bol}}+\mu=B+BC_B$, the resulting semi-empirical $V$ band P-L relation is: $V=-2.723(\pm0.030)\log P + 17.050(\pm0.023)$ with a dispersion of $0.219$. Doing the same for the $B$ band data, the semi-empirical $B$ band P-L relation is: $B=-2.361(\pm0.041)\log P + 17.345(\pm0.031)$ with a dispersion of $0.298$. Both of the semi-empirical P-L relations are in good agreement with equation (1) \& (2): the difference in the slope is less than $0.010$mag. and the zero-points are almost identical in both passbands. For testing the semi-empirical $I$ band P-L relation, we use both of the $B$ and $V$ band mean magnitudes and $M_{\mathrm{bol}}+\mu=[(B+BC_B)+(V+BC_V)]/2$ for the common Cepheids to derive equation (3), and the resulted semi-empirical $I$ band P-L relation is: $I=-2.979(\pm0.023)\log P + 16.599(\pm0.018)$ with a dispersion of $0.164$, which is still in good agreement with equation (3). Therefore, the above semi-empirical approach can be used to derive P-L relations in Sloan magnitudes, which we summarize in Table \ref{tab2}. The plots of these P-L relations are presented in Figure \ref{pl}. Note that the dispersion of the P-L relation decreases from $u$ to $z$ passbands. This reduction of the dispersion as wavelength increases is well-known in the Cepheid community \citep[see, for example,][]{mad91}. 

\begin{deluxetable}{lccc}
\tabletypesize{\scriptsize}
\tablecaption{Semi-empirical $ugriz$ P-L relations.\label{tab2}}
\tablewidth{0pt}
\tablehead{
\colhead{$\lambda$} & 
\colhead{Slope} & 
\colhead{Zero-Point} & 
\colhead{$\sigma$} }
\startdata
$u$  & $-1.981\pm0.046$ & $18.195\pm0.035$ & 0.338 \\ 
$g$  & $-2.518\pm0.036$ & $17.165\pm0.027$ & 0.262 \\ 
$r$  & $-2.819\pm0.027$ & $17.027\pm0.020$ & 0.193 \\ 
$i$  & $-2.928\pm0.023$ & $17.032\pm0.018$ & 0.171 \\ 
$z$  & $-3.007\pm0.022$ & $17.064\pm0.017$ & 0.160 
\enddata
\tablecomments{$\sigma$ is the dispersion of the P-L relation.}
\end{deluxetable}

\section{A Sanity Check: Empirical P-L Relations from Photometric Transformations}

The Johnson-Sloan photometric transformations that are available in the literature can be used to transform the $BV$ mean magnitudes for our LMC Cepheids to the Sloan magnitudes. The transformation for $u'g'r'i'z'$ magnitudes (from the United States Naval Observatory 40-inch telescope) are availale, for example, in \citet{fuk96}, \citet{smi02} and \citet{rod06}; while the transformation for $ugriz$ magnitudes (from the SDSS 2.5-meter telescope) are given in \citet{bil05}, \citet{jes05} and \citet{jor06}. These transformations in general take the form of $X-V=a_0(B-V)+a_1$, where $X=\{g',r',g,r\}$, for the $(B-V)$ color. Therefore the observed $V$ band mean magnitudes and $(B-V)$ colors for the LMC Cepheids can be transformed to the Sloan magnitudes, and the P-L relations can be fitted in these passbands. This serves as an independent check to our results presented in the previous section. In this paper we apply the transformation from \citet{jes05} only, but the resulting P-L relations from other transformations are available upon request. Using \citet{jes05} transformation, we obtain: $g=-2.514(\pm0.036)\log P+17.119(\pm0.028)$ with a dispersion of $0.265$ and $r=-2.855(\pm0.027)\log P+17.025(\pm0.020)$ with a dispersion of $0.195$, which are in good agreement to the P-L relations given in Table \ref{tab2}. 

However, there is a major drawback to using the Johnson-Sloan photometric transformation to obtain the empirical P-L relations. The empirical Johnson-Sloan photometric transformations available in the literature are mostly derived from the standard stars, which could span a wide range of luminosity class \citep{smi02,rod06}, effective temperature and/or metallicity \citep[which could be close to the Solar value,][]{jor06}. On the other hand, Cepheids are cool supergiants with spectral type from $F$ to $K$. Therefore the photometric transformations may not be applicable to the LMC Cepheids\footnote{We thank J. Hartman \& D. Bersier for pointing this out.}. To test this, we use the same theoretical bolometric corrections in Section 2 to obtain the theoretical Johnson-Sloan transformation for two ``extreme'' cases: 

\begin{figure}
\plotone{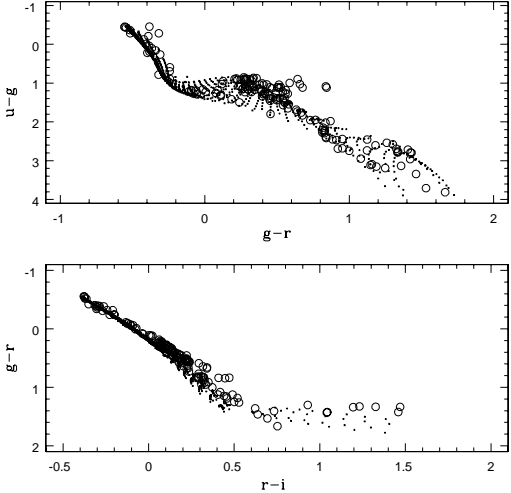}
\caption{Comparison of the colors from grids of theoretical calculations (small crosses) to the observation of standard stars taken from \citet[][open circles, after converting the $u'g'r'i'z'$ colors to $ugriz$ colors using the transformation given in SDSS website]{smi02}. \label{sdss}}
\end{figure}

\begin{description}
\item Case 1: We use the grids of $BC$ with $4000\mathrm{K} \leq T_{eff} \leq 7000\mathrm{K}$ and $0.0 \leq \log (g) \leq 2.5$, which are appropriate for Cepheids at mean light with metallicity $[M/H]=-0.3$, to represent the LMC Cepheids. 

\item Case 2: We use the grids of $BC$ with $3100\mathrm{K} \leq T_{eff} \leq 80000\mathrm{K}$ and $0.0 \leq \log (g) \leq 5.0$, to represent the standard stars that cover a wide range of effective temperature and surface gravity, and metallicity of $[M/H]=0.0$ and $[M/H]=-0.5$ which roughly bracket the Solar metallicity and LMC-type metallicity. Figure \ref{sdss} compares the grids of theoretical colors with the colors of standard stars, which shows that the theoretical colors are well covered by the observation of standard stars.
\end{description}

\ni   The Johnson-Sloan transformations are then derived from these grids of theoretical colors. After obtaining the transformations for the two cases as mentioned, we transform our LMC Cepheid data to the Sloan magnitudes and fit the linear P-L relations. The differences ($\Delta$) of the slope from the fitted $gr$ P-L relations for these two cases are: $\Delta g=0.011\pm0.052$ and $\Delta r=0.016\pm0.038$; while the difference for the zero-point are: $\Delta g=0.001\pm0.040$ and $\Delta r=0.009\pm0.029$. The small differences between the P-L relations of Case 1 and Case 2 implies that the derived P-L relation in the Sloan magnitudes, at least in the $gr$ passbands, will be insensitive to the adopted transformations. Therefore, we believe the adopted transformations from the literature should not significantly affect the empirical P-L relations. The work for verifying the transformations for Cepheids in a more proper and rigorous way is currently underway with Cepheids in M33 by Bersier et al. (in preparation -- private communication).

\section{Conclusion}

Using the observed $BVI$ mean magnitudes from LMC Cepheids and the theoretical bolometric corrections from Padova group, we derive semi-empirical Cepheid P-L relations in the Sloan magnitudes that can be used in the current and future distance scale and Cepheid works. For a sanity check, we compare the $gr$ band P-L relations derived from adopting the available Johnson-Sloan photometric transformation in the literature to our semi-empirical P-L relations. The resulting comparison finds that these two sets of P-L relation are generally consistent with each other. 

\acknowledgements
We thank the anonymous referee for helpful suggestions to improve the manuscript. We also thank J. Hartman \& D. Bersier for useful discussion and suggestions. SMK acknowledges the support from HST-AR-10673.04-A and the Chretien International Research Award of the American Astronomical Society. CN acknowledges support from NSF award OPP-0130612 and a University of Illinois seed funding award to the Dark Energy Survey. This research was supported in part by NASA through the American Astronomical Society's Small Research Grant Program.

\end{document}